\documentclass[aps,pra,twocolumn,nofootinbib,superscriptaddress]{revtex4}
\pdfoutput=1
\usepackage{epsfig}
\usepackage{amsmath}
\usepackage{multirow}
\usepackage{blindtext}
\usepackage{hyperref}
\usepackage{amssymb}
\usepackage{ragged2e}
\usepackage{threeparttable,booktabs}
\usepackage{diagbox}
\usepackage{hhline}
\usepackage{color}
\usepackage{bm}
\usepackage{xfrac}
\usepackage{anyfontsize}

\def\etal{{\it et al}}

\newcommand{\HNAS}{
Institute of Quantum Materials and Physics, Henan Academy of Sciences, Zhengzhou 450046, China\\
}

\newcommand{\RUG}{
Van Swinderen Institute for Particle Physics and Gravity,
University of Groningen, Nijenborgh 4, 9747 Groningen, The Netherlands\\
}

\newcommand{\JAEA}{
Advanced Science Research Center, Japan Atomic Energy Agency (JAEA), Tokai, Ibaraki 319-1195, Japan\\
}

\newcommand{\TAU}{
School of Chemistry, Tel Aviv University, 6997801 Tel Aviv, Israel\\
}

\newcommand{\CU}{
Department of Physical and Theoretical Chemistry \& Laboratory for Advanced Materials, Faculty of Natural Sciences, 
Comenius University, Mlynsk\'{a} dolina, 84215 Bratislava, Slovakia\\
}
\usepackage{graphicx}
\usepackage{dcolumn}
\usepackage{bm}

\usepackage[utf8]{inputenc}
\begin{document}
\preprint{APS/123-QED}
\title{Relativistic coupled cluster calculations of the electron affinity and ionization potentials of lawrencium}
\author{Yangyang Guo}
\email{gyangyang@wustl.edu}
\affiliation{\HNAS}
\affiliation{\RUG}

\author{Luk\'{a}\v{s} F. Pa\v{s}teka}
\affiliation{\CU}
\affiliation{\RUG}

\author{Yuichiro Nagame}
\affiliation{\JAEA}

\author{Tetsuya K. Sato}
\affiliation{\JAEA}

\author{Ephraim Eliav}
\affiliation{\TAU}

\author{Marten L. Reitsma}
\affiliation{\RUG}

\author{Anastasia Borschevsky}
\email{a.borschevsky@rug.nl}
\affiliation{\RUG}
\begin{abstract}
 The calculations of the first and the second ionization potentials of lawrencium and lutetium and the electron affinity of lawrencium are performed within the relativistic coupled cluster framework. These results are corrected by including the contributions of extrapolation to the complete basis set limit and higher order contributions due to relativity and electron correlation. The excellent agreement between our predictions of the ionization potentials of Lu and Lr and experimental values supports the accuracy of our predictions of the second ionization potential and the electron affinity of Lr.
 \end{abstract}
\maketitle

\section{Introduction}
The discovery of superheavy element oganesson ($Z$~=~118) completes the 7th row of the periodic table \cite{oganessian2007heaviest,karol2016discovery,oganessian2015super}. At the same time, atomic spectroscopy of the heaviest elements is notoriously challenging, and our experimental knowledge of electronic structure presently terminates at the later actinides.

Actinides were investigated via a variety of experimental approaches, such as laser spectroscopy, ion chemistry, and surface ionization techniques, see Refs. \cite{doi:10.1021/cr3002438, NAGAME2015614, sato2015measurement, doi:10.1021/jacs.8b09068, PhysRevLett.120.263003}. Such studies allow the exploration of the fascinating behaviour of these heavy species, where the relativistic effects play a major role \cite{pershina2019relativity}. 


At the very end of the actinide series, lawrencium ($Z=103$) was synthesized for the first time in 1961 by bombardment of californium with boron ions \cite{PhysRevLett.6.473}. 
Due to the strong relativistic effects that stabilize the outermost 7p$_{\sfrac{1}{2}}$ orbital, the electronic configuration of the ground state of Lr is predicted to be [Rn]5f$^{14}$7s$^2$7p$_{\sfrac{1}{2}}$ \cite{refId0-Lr,PhysRevA.90.012504, borschevsky2007transition, PhysRevLett.88.183001, PhysRevA.52.291, KahRaeEli21}, in contrast to its lighter homologue Lu, [Xe]4f$^{14}$6s$^2$5d. 
The ionization potential (IP$_1$) of Lr was first measured on an atom-at-a-time scale using the surface ionization technique by Sato et al. \cite{sato2015measurement,doi:10.1021/jacs.8b09068}. These measurements were accompanied by the state-of-the-art relativistic calculations. 
Recently the lower limit on the second ionization potential (IP$_{2}$) of Lr was established in a gas-phase ion chemistry experiment \cite{doi:10.1021/acs.jpca.1c01961}. 
Electron affinity (EA) is another fundamental electronic property that determines chemical behaviour and reactivity. So far, the electron affinity of Lr has not been measured, but the development of novel, specially designed experimental techniques brings this goal closer to our reach.

Recently, the EA of the radioactive astatine (At, atomic number $Z = 85$) isotope, $^{211}$At (half-life, $T_{\sfrac{1}{2}}$~=~7.2~h) produced in a spallation reaction of thorium (Th) with 1.4~GeV protons at the CERN-ISOLDE radioactive ion beam facility, was successfully measured by laser photodetachment threshold spectroscopy \cite{leimbach2020electron}. In this innovative experiment, $3.75 \times 10^{6}$ particles per second of $^{211}$At$^{-}$ ions were produced. The $^{211}$At isotope was mass-separated by ISOL (Isotope Separator On-Line) and guided into the GANDALPH (Gothenburg ANion Detector for Affinity measurements by Laser PHotodetachment) apparatus which is designed for measurements of the EAs of radioactive elements \cite{rothe2017laser}. The EA of astatine was determined to be 2.41578(7)~eV, in excellent agreement with the accompanying state-of-the-art relativistic coupled cluster calculations that predicted an EA of 2.414(16)~eV.

However, measurement of EA of still heavier elements, especially those heavier than fermium (Fm, $Z = 100$), is an extremely challenging task since these elements must be produced using accelerator-based heavy-ion-induced nuclear reactions \cite{,oganessian1975experiments,zagrebaev2014gas}. Production rates of these elements, thus, are minuscule and the half-lives of the produced isotopes are very short; they are usually available only in very small quantities that require one-atom-at-a-time scale experiments and measurements in the approximate half-life range of the isotopes under investigation \cite{single.atom,doi:10.1021/cr3002438,sch15}.

The isotopes $^{256}$Lr (T$_{\sfrac{1}{2}} = 27$~s) and $^{255}$Lr (T$_{\sfrac{1}{2}} = 31$~s) with sufficiently long half-lives, produced at rates of a few atoms per second, will be possible candidates for the EA measurements of Lr. $^{256}$Lr produced in the bombardment of a $^{249}$Cf target with $^{11}$B beams was exploited for the measurement of the first ionization potential of Lr. In the experiment, the Lr atoms were extracted as Lr$^{+}$ ions using ISOL at JAEA (Japan Atomic Energy Agency) \cite{sato2015measurement}. $^{255,256}$Lr synthesized in the reactions of $^{48}$Ca ions with a $^{209}$Bi target were employed in the direct mass measurements of these nuclides using the Penning trap mass spectrometer SHIPTRAP at GSI (GSI Helmholzzentrum für Schwerionenforshung) \cite{ramirez2012direct}. 

For the measurement of the electron affinity of Lr, effective production of negative Lr$^{-}$ ions is indispensable. This can be achieved using strong electron donors, such as cesium, with low IP$_{1}$. A Cs-sputter negative ion-source was applied in the past to produce negative ions of lutetium (Lu$^{-}$), lanthanide homologue of Lr \cite{davis2001measurement}. 
Similarly, the production of Lr$^-$ ions could be achieved via the charge exchange reaction Lr$^{+}$ → Lr$^{-}$ by passing the Lr$^{+}$ ion-beams through Cs vapor.
While the measurement of EA is rendered challenging by the difficulty in obtaining Lr atoms on the scale of more than one atom at a time, an efficient detection system of radioactive $\alpha$-decays of Lr isotopes event by event could provide a significant advantage for such measurements \cite{khuyagbaatar201448+}.
Efficient production of negative ions, coupled to laser photodetachment spectrometry and $\alpha$-decay spectrometry could thus open up the possibility of measuring electron affinities of heavy elements, including Lr. Another component crucial for the success of these challenging experiments on short-lived and rare species are accurate and reliable theoretical predictions of the EA of Lr.

The present work aims to provide benchmark values of the first and second IPs and EAs of Lr, using the state-of-the-art relativistic coupled-cluster approach. An extensive computational study is used to estimate the uncertainties of our predictions, following the scheme developed in our earlier works \cite{GuoPasEli21,leimbach2020electron,guo2022relativistic}. 
Alongside the calculations for Lr, we also carried out equivalent investigations of the first and second IPs of its lighter homologue, Lu. The accuracy of our predictions for the EA and the IP$_2$ of Lr is confirmed by comparison of both IPs of Lu and the IP$_{1}$ of Lr to the experimental values \cite{refId0-Lu,maeda1989highly,PhysRevA.95.052501,sato2015measurement}. 

Several earlier accurate calculations of atomic properties of Lr are available, based on different approaches. Transition energies and ionization potentials were calculated using the Fock-space coupled cluster (FSCC) method \cite{PhysRevA.52.291,borschevsky2007transition,Kahl_2019}, the configuration interaction approach combined with many-body perturbation theory (CI+MBPT) \cite{kahl2021ab}, and the combination of the configuration interaction method and all-order single-double coupled cluster technique (CI+all-order); in the latter work also static dipole polarizabilities were presented \cite{PhysRevA.90.012504}.   The only EA prediction so far was carried out within the FSCC method \cite{PhysRevA.52.291,borschevsky2007transition}. A higher accuracy prediction of this property, along with reliable error bars is important for providing solid support for future experiments.\\
\indent

\section{METHOD AND COMPUTATIONAL DETAILS}

The overall computational methodology and the scheme for the evaluation of the uncertainties is similar to that used in our recent works (see Refs. \cite{GuoPasEli21,leimbach2020electron,guo2022relativistic}). 
The calculations were carried out in the framework of the relativistic single-reference coupled-cluster approach with single, double, and perturbative triple excitations (DC-CCSD(T)). The Dirac--Coulomb (DC) Hamiltonian is (in atomic units), 
\begin{eqnarray}
H_\text{DC}= \displaystyle\sum\limits_{i}h_\text{D}(i)+\displaystyle\sum\limits_{i<j}(1/r_{ij}),
\label{eqHdcb}
\end{eqnarray}
where $h_\text{D}$ is the relativistic one-electron Dirac Hamiltonian,
\begin{eqnarray}
h_\text{D}(i)=c\bm{\alpha }_{i}\cdot \mathbf{p}_{i}+c^{2}\beta _{i}+V^n(i),
\label{eqHd}
\end{eqnarray}
and $\alpha$ and $\beta$ are the four-dimensional Dirac matrices. The nuclear potential $V^n(i)$ is modelled by a Gaussian charge distribution \cite{VisDya97}.

The DC-CCSD(T) calculations were carried out using the DIRAC19 computational program package \cite{DIRAC19}. In the coupled-cluster calculations, all electrons were correlated and virtual orbitals with energies above 70 a.u. were omitted. We have carried out test calculations with a higher active-space cutoff and used the results in the error estimate (see below).

All calculations were performed using Dyall’s relativistic basis sets \cite{Dyall2007, Gomes2010, Dyall2012}, consisting of uncontracted Gaussian functions, namely the valence (v$N$z), the core-valence (cv$N$z) and the all-electron (ae$N$z) basis sets, with $N$ the basis set cardinality, $N=2,3,4$. We found in our previous studies that diffuse functions of different angular momenta have a significant effect on the calculated atomic properties, especially on the electron affinity \cite{GuoPasEli21}. Thus the basis sets were augmented with diffuse functions until convergence of the calculated IPs and EA was achieved. These augmented basis sets are designated ($x$-aug)-cv$N$z, where the prefix $x$ stands for the number of layers of diffuse functions added to the basis set.

We performed extrapolations to the complete basis set (CBS) limit, using the scheme of Feller~\etal~\cite{Feller1992} for the DHF values (3-point extrapolation using 2z, 3z and 4z results) and the scheme of Helgaker \cite{HelKloKoc97} for the correlation contribution (2-point extrapolation using the 3z and the 4z results).\

In the DC Hamiltonian, the electronic repulsion is taken in its non-relativistic form. Due to the non-instantaneous interaction between particles being limited by the speed of light in the relativistic framework, a correction to the two-electron part of $H_\text{DC}$ is added (in Coulomb gauge), in the form of the zero-frequency Breit interaction:
\begin{eqnarray}
B_{ij}=-\frac{1}{2r_{ij}}[\bm{\alpha }_{i}\cdot \bm{\alpha }_{j}+(%
\bm{\alpha }_{i}\cdot \mathbf{r}_{ij})(\bm{\alpha }_{j}\cdot \mathbf{%
r}_{ij})/r_{ij}^{2}].
\label{eqBij}
\end{eqnarray}

To account for the QED corrections, and thus further improve the accuracy of our results, we applied the model Lamb shift operator (MLSO) of Shabaev and co-workers \cite{ShaTupYer15} to the atomic no-virtual-pair many-body Dirac-Coulomb-Breit (DCB) Hamiltonian as implemented into the QEDMOD program. This model Hamiltonian uses the Uehling potential and an approximate Wichmann--Kroll term for the vacuum polarization (VP) potential \cite{BLOMQVIST197295} and local and nonlocal operators for the self-energy (SE), the cross terms (SEVP), and the higher order QED terms \cite{PhysRevA.88.012513}. The implementation of the MLSO formalism in the Tel Aviv atomic computational package \cite{TRAFS-3C} allows us to obtain the VP and SE contributions beyond the usual mean-field level, namely at the FSCC level. The FSCC methodology within the same code was also used to calculate the Breit interaction. 
These contributions were then added to the DIRAC19 CCSD(T) IPs and EAs and the resulting values are designated DCB-CCSD(T) and DCB-CCSD(T)+QED.


We improved the accuracy of our calculations further by going beyond the standard CCSD(T) approach and considering higher-order excitations. The full triple contributions $\Delta$T were calculated with the EXP-T program \cite{Oleynichenko2020}. Dyall's (2-aug)-v3z basis sets were used for the calculation of $\Delta$T. For Lr the valence 7s and 7p and core 5d, 5f, 6s, 6p electrons were correlated, while the virtual space comprised of 99 orbitals (up to 70 a.u.). For Lu the valence 6s and 5d and core 4-5s, 4-5p, 4f electrons were correlated, while the virtual space comprised of 94 orbitals (up to 70 a.u.). We have found in our earlier investigations that higher-order excitations are generally localised in the valence-shell region \cite{PasEliBor17}, justifying our use of a limited active space. The differences between the full triples, and the perturbative triples ($\Delta\text{T}=\text{T}-\text{(T)}$) were added to the DCB-CCSD(T)+QED results (extrapolated to the CBS limit) to obtain the final recommended values, DCB-CCSDT+QED. 





\section{RESULTS AND DISCUSSION}
The calculated IP$_{1}$, IP$_{2}$, and EA of Lr and IP$_{1}$ and IP$_{2}$ of Lu are presented in Table \ref{tab:basis set} for different basis sets; these calculations were carried out on the DC-CCSD(T) level of theory. 
The first three lines explore the effect of the core-correlating functions, using the 4z quality basis sets. Since in our calculations, all the electrons are correlated, such functions have a significant effect on the quality of the results. The difference between cv4z and v4z ionization potentials is especially notable, for EAs the effect is less significant. Switching to the all-electron (ae4z) basis set yields negligible changes in the calculated properties, with the exception of IP$_{2}$ of Lr, where it increases the calculated value by 6 meV, representing a mere 0.04\% difference. Thus, we proceed with further calculations using the cv4z basis sets and accounting for this difference in the estimated uncertainty. 

The ionization potentials of both Lr and Lu are not sensitive to the addition of diffuse functions,  while for the EA of Lr we observe an increase of 2\% upon the addition of the first augmentation layer ((1-aug)-cv4z). This is expected, the electron affinity represents the energy associated with a loosely bound extra electron, rendering the quality of the description of the outer part of the wavefunction important.  Adding two further augmentation layers has a small effect of 2 meV the obtained electron affinity ((3-aug)-cv4z). We thus use the (3-aug)-cv$N$z basis set family for the extrapolation of the results to the complete-basis-set limit ((3-aug)-CBS-cv$N$z). 
\begin{table}[h!]
  \centering
    \caption{IP$_{1}$, IP$_{2}$ and EA(eV) of Lr and IP$_{1}$ and IP$_{2}$ of Lu on the DC-CCSD(T) level of theory using different quality of basis sets.}
    \begin{tabular}{@{\extracolsep{4pt}}l c c c c c@{}}
    \hline\hline

\multirow{2}{*}{Basis set}&\multicolumn{2}{c}{Lu}& \multicolumn{3}{c}{Lr}\\
    \cline{2-3}\cline{4-6}
    &IP$_1$&IP$_2$&IP$_1$&IP$_2$&EA\\
   \hline
v4z&5.305&13.962&4.943&	14.502&	0.424\\
cv4z 
&5.332&13.988&4.947&14.536&0.421\\ 
ae4z &5.332&13.988&4.947&14.542&0.420\\
 (1-aug)-cv4z&5.333&13.988&4.947&14.536&0.428\\
(3-aug)-cv2z& 5.259&13.945&4.908&14.475&0.389\\
(3-aug)-cv3z& 5.308&13.979&4.924&14.481&0.408\\
  (3-aug)-cv4z& 5.336&13.988&4.947&14.534&0.426\\
  (3-aug)-CBS-cv$N$z &5.356&13.995&4.963&14.573&0.439\\
     \hline\hline
  \end{tabular}
  \label{tab:basis set}
\end{table}

\begin{table}[h!]
  \centering
    \caption{IP$_{1}$, IP$_{2}$ and EA(eV) of Lr and IP$_{1}$ and IP$_{2}$ of Lu with higher order corrections contributions (eV).}
    \begin{tabular}{@{\extracolsep{4pt}}l c c c c c@{}}
    \hline\hline

\multirow{2}{*}{Basis set}&\multicolumn{2}{c}{Lu}& \multicolumn{3}{c}{Lr}\\
    \cline{2-3}\cline{4-6}
    &IP$_1$&IP$_2$&IP$_1$&IP$_2$&EA\\
   \hline
CCSD&5.260&13.831&4.903&14.394&0.329\\
CCSD(T)&5.356&13.995&4.963&14.573&0.439\\
CCSDT&5.370&13.995	&4.961	&14.574&0.448\\
CCSDT+Breit&5.376&13.987&4.949&14.558&0.447\\
CCSDT+Breit+QED&5.387&14.023&4.954&14.618&0.446\\


     \hline\hline
  \end{tabular}
  \label{tab:higher order}
\end{table}

Table \ref{tab:higher order} summarizes the results including higher-order correction contributions. Among them, the perturbative triple contributions are between 60 and 170 meV, making it the largest term. In particular, in the case of the electron affinity of Lr, perturbative triples comprise about a quarter of the total value. Transitioning from perturbative to full triple excitations increases the IP$_{1}$ of Lu by 15 meV and the EA of of Lr by 9 meV, respectively, while the changes for the other properties remain within 2 meV. The Breit correction increases the IP$_{1}$ of Lu by 6 meV while decreasing the IP$_{2}$ of Lu and the IPs of Lr by 8 to 16 meV. QED corrections notably boost the IPs of both Lu and Lr, particularly elevating the IP$_{2}$ of Lr by 60 meV. Both Breit and QED corrections have a negligible effect on the EA of Lr. The significant QED contribution to the IPs, compared with the size of the Breit effect, should stimulate further investigation into the QED effects in heavy elements \cite{malyshev2022model}.


Several sources of uncertainty arise from the approximation inherent to the computational approach and from the computational limitations; these are the incompleteness of the basis set, the neglect of electron correlation beyond triple excitation, the restriction of the correlation space and the neglect of higher-order QED corrections. 

The uncertainty stemming from the basis set arises from three main factors: the extrapolation to a complete basis set using a semi-empirical scheme, the limited augmentation of the basis set, and the absence of inner layer electron description in the core-valence basis set. The presented calculations of the last line in Table \ref{tab:basis set} are performed with the Helgaker scheme; here we compare the Helgaker, Lesiuk, and Martin schemes \cite{HelKloKoc97,doi:10.1021/acs.jctc.9b00705,MARTIN1996669,MARTIN2006} for extrapolation, respectively, as detailed in \cite{guo2022relativistic} and determine the 95\% confidence interval of the standard deviation among the three schemes to represent the complete basis set (CBS) uncertainty, as shown in Table \ref{tab:scheme}. Additionally, we consider the difference between the (3-aug)-cv4z and (1-aug)-cv4z results as the augmentation uncertainty and the difference between the values obtained with the ae4z and cv4z basis sets as uncertainty due to basis set incompleteness in the core region. The latter is also added as a correction to the final recommended values in Table~\ref{summary}.

The uncertainty arising from the treatment of electron correlation is due to the cutoff of the virtual correlation space (we correlate all the electrons) and the neglect of higher-order excitations beyond CCSDT. To verify the effect of energy cutoff in virtual space, we increased the cutoff from 70~a.u. to 2000~a.u. using the v4z basis set and took the difference as the corresponding uncertainty. This correction was also added to the final recommended values in Table~\ref{summary}. The neglected higher-order excitations beyond the triple level were estimated to be no greater than $\Delta$T. 

Regarding relativity, we assume that the neglected higher-order QED contributions are not larger than the second order contributions.
Based on the expansion of the bound-state propagator \cite{SunSalSau22}, the leading order and next-to-leading order QED contributions scale as $\sim Z \alpha^2$ and $\sim Z^2 \alpha^3$, respectively, where $\alpha$ is the fine structure constant. We thus used the calculated QED contributions scaled with the ratio $Z \alpha$ as the corresponding uncertainties.

The individual contributions to the uncertainty are detailed in Table \ref{tab:error}. One can assume that the uncertainty contributions stemming from different sources are independent to a large degree; this assumption should be valid as long as we are dealing with higher-order contributions. Thus the total uncertainty is obtained by adding the individual sources of uncertainty using the usual Euclidean norm.

\begin{table}[t]
  \centering
    \caption{First and second IPs and EA of Lr and IP$_{1}$, IP$_{2}$ of Lu obtained using different extrapolation schemes together with the resulting 95\% confidence interval, in eV.}
    \begin{tabular}{@{\extracolsep{4pt}}llllll@{}}
    \hline\hline
    \multirow{2}{*}{Scheme}&\multicolumn{2}{c}{Lu}& \multicolumn{3}{c}{Lr}\\ \cline{2-3}\cline{4-6}
    & \multicolumn{1}{c}{IP$_{1}$} & \multicolumn{1}{c}{IP$_{2}$} & \multicolumn{1}{c}{IP$_{1}$}  & \multicolumn{1}{c}{IP$_{2}$} & \multicolumn{1}{c}{EA} \\
   \hline
   Helgaker&	5.356	&13.995&4.963&	14.573	&0.439\\
Lesiuk&5.361	&13.997&4.967&	14.583	&0.443	\\
Martin&5.351	&13.993&	4.960	&14.565&	0.437	\\
95\% c.i.&0.009&\phantom{1}0.004&0.007&\phantom{1}0.017&0.006\\ 
   \hline\hline
   \end{tabular}  
   \label{tab:scheme}
\end{table}

\begin{table}[t]
 \centering
    \caption{Main sources of uncertainty in the calculated IP$_{1}$, IP$_{2}$ and EA of Lr and IP$_{1}$ and IP$_{2}$ of Lu, in meV.}
    \begin{tabular}{@{\extracolsep{5pt}}l  r r r r r@{}}
    \hline\hline
    \multirow{2}{*}{Error source} &\multicolumn{2}{c}{Lu}& \multicolumn{3}{c}{Lr}\\ \cline{2-3}\cline{4-6}
    & \multicolumn{1}{c}{IP$_{1}$} & \multicolumn{1}{c}{IP$_{2}$} & \multicolumn{1}{c}{IP$_{1}$}  & \multicolumn{1}{c}{IP$_{2}$} & \multicolumn{1}{c}{EA} \\
    \hline

  Basis set\\
  \quad-- CBS scheme &9.4&3.8&7.2&17.4&6.1  \\ 
  \quad-- augmentation &3.4&0.1&0.0&--0.5&--1.8\\
  \quad-- (ae4z -- cv4z) &--0.2&0.0&0.6&5.7&--0.5\\
  Correlation\\
  \quad-- virtual cutoff &4.0&3.0&0.9&3.5&--0.3\\
  \quad-- higher excitations &13.8&0.3&--2.2&0.8&9.0\\
  QED & 5.7 & 18.7 & 3.8 & 45.0 & --0.8\\
  
  Total &18\phantom{.0}&19\phantom{.0}&9\phantom{.0}&49\phantom{.0}&11\phantom{.0} \\
      \hline\hline
  \end{tabular}
    \label{tab:error}
\end{table}

\begin{table*}[!tp]
  \centering
    \caption{Final recommended values of the IPs(eV) and EA(eV) of Lr and Lu compared with previous calculations and experiment, where available.}
    \begin{tabular}{@{\extracolsep{4pt}}l lllllll @{}}
    \hline\hline
    \multirow{2}{*}{Method}&\multirow{2}{*}{Year}&\multicolumn{2}{c}{Lu}& \multicolumn{3}{c}{Lr}&\multirow{2}{*}{Reference} \\
    \cline{3-4}\cline{5-7}
    &&\multicolumn{1}{c}{IP$_{1}$($6s^2$)}&\multicolumn{1}{c}{IP$_{2}$($6s^1$)}&\multicolumn{1}{c}{IP$_{1}$($7s^2$)}&\multicolumn{1}{c}{IP$_{2}$($7s^1$)}&\multicolumn{1}{c}{EA($7s^27p^2$)}&\\
   \hline
    
   \hline
   CBS-CCSDT+Breit+QED&&5.391(18)&14.026(19)&4.955(9)&\phantom{$>$}14.627(49)&0.446(11)&Present\\
FSCC&2007&5.301&&4.894&&0.476&\cite{borschevsky2007transition}\\
CI+all order&2014&&&	4.934&&&\cite{PhysRevA.90.012504}\\
CCSD(T)&2015&&&4.963(15)&&&\cite{sato2015measurement}\\
FSCC&2021&&13.973&&\phantom{$>$}14.500(48)&&\cite{Kahl_2019}\\
Experiment&&5.4259(13) &14.13(10)\makeatletter\def\Hy@Warning#1{}\makeatother\footnote{Indirectly derived from experimental data in the lanthanide series.}  &$4.96^{+0.05}_{-0.04}$&$>$13.3(3)\makeatletter\def\Hy@Warning#1{}\makeatother\footnote{Only lower limit is predicted.}&&\cite{maeda1989highly,PhysRevA.95.052501,johnson2017lanthanide,sato2015measurement,doi:10.1021/jacs.8b09068,doi:10.1021/acs.jpca.1c01961}\\
     \hline\hline
  \end{tabular}
  \label{summary}
\end{table*}
The final recommended values of the first and second IPs of Lu and Lr, as well as the EA of Lr are summarized in Table \ref{summary}, along with recent-high accuracy calculations and the existing experimental values. \\

Our calculated IP$_{1}$ of Lu is 34 meV below the experimental value, with this difference exceeding the uncertainty we set on our prediction. It is not clear whether the source of this discrepancy is in the computational scheme or whether it is experimental in origin.
Our prediction of IP$_{2}$ of Lu agrees well with the recently proposed semiempirical values \cite{johnson2017lanthanide} and within the combined uncertainties of the two results.
The present IP$_{1}$ value for Lr stands at 4.955 eV with an uncertainty of $\pm$ 9 meV, in good agreement with the earlier CCSD(T) prediction from 2015. 
 Remarkably, both these findings align well with experimental data, which reports an IP$_{1}$ of $4.96^{+0.05}_{-0.04}$ eV \cite{sato2015measurement}. These values are also in good agreement with the prediction obtained using the CI+all order technique \cite{PhysRevA.90.012504}. The earlier result obtained using the FSCC method is approximately 0.07 eV lower than the experimental value. This discrepancy can be attributed to the omission of electron correlation beyond double excitations. 

Utilizing the one-atom-at-a-time gas-phase ion chemistry technique, the lower limit on the IP$_{2}$ of Lr was set at 13.3(3) eV \cite{doi:10.1021/acs.jpca.1c01961}. The current predictions surpass this value by more than an eV, and are in good agreement with the earlier FSCC value \cite{Kahl_2019}. 

Our prediction for the EA of Lr is in good agreement with the 2007 FSCC result, but expected to be more accurate due to the basis set extrapolation and to the inclusion of higher excitations and QED corrections. The proposed error bars on this value should be used with caution however, due to the possible underestimation of uncertainty for the IP$_{1}$ of Lu. 

\section{CONCLUSION\label{IV}}
We carried out relativistic coupled-cluster calculations of the first and second ionization potentials and electron affinity of lawrencium, as well as the first and second ionization potentials of lutetium, the latter allowing comparison with experimental data. We have extrapolated our results to the complete basis set limit and corrected the calculated properties for higher-order correlation effects and for the Breit and QED contributions. 

Our calculations are in good agreement with the available experimental values for the first ionization potential of Lr and the second ionization potential of Lu, confirming the accuracy and reliability of the selected computational approach. Future efforts aimed at determining the electron affinity and the second ionization potential of Lr will benefit from our theoretical prediction, which includes uncertainty estimation, providing a detection range to guide the measurements.


\section*{Acknowledgments}
We would like to thank the Center for Information Technology of the University of Groningen for their support and for providing access to the Peregrine and Hábrók high-performance computing clusters.
LFP acknowledges the support from the project number VI.C.212.016 of the talent programme VICI financed by the Dutch Research Council (NWO), and from the Slovak Research and Development Agency projects APVV-20-0098 and APVV-20-0127.

\bibliographystyle{naturemag}
\bibliography{bib}

\end{document}